\documentclass[12pt]{article}

\title{Extending The Range of Application of Permutation Tests: the Expected Permutation p-value Approach} 
\author{  Daniel Commenges \\   INSERM U897, ISPED, 146 rue L\'eo Saignat, Bordeaux, 33076, France\\Tel: (33) 5 57 57 11 82; Fax (33) 5 56 24  00 81}

\begin{document}

\maketitle

{\bf Abstract}

 The limitation of permutation  tests is that they assume exchangeability.  It is shown that in generalized linear models one can construct permutation tests from score statistics in particular cases. When under the null hypothesis the observations are not exchangeable, a representation in terms of Cox-Snell residuals allows to develop an approach based on an expected permutation p-value (Eppv); this is applied to the logistic regression model. A small simulation stydy and an illustration with real data are given.

{\bf Resum\'e}

 La limitation des tests de permutation est qu'ils sont bas\'es sur une hypoth\`ese d'\'echangeabilit\'e. Il est montr\'e que dans les mod\`eles lin\'eaires g\'en\'eralis\'es on peut construire des tests de permutation par la statistique du score dans des cas particuliers. Quand les observations ne sont pas \'echangeables sous l'hypoth\`ese nulle, une repr\'esentation en terme de r\'esidus de Cox-Snell permet de d\'evelopper une approche bas\'ee sur l'esp\'erance de la p-valeur de permutation; ceci est appliqu\'e au mod\`ele de r\'egression logistique.


{\bf Version fran\c caise abr\'eg\'ee}

Consid\'erons une statistique $T(Y)$ pour tester une  hypoth\`ese $H_0$.
La d\'ecision de  rejet de $H_0$ est prise si $T(Y) \ge c_{\alpha}$,
$c_{\alpha}$ choisi tel que l'erreur de  type I est  $\alpha$.
La p-valeur est d\'efinie comme une variable al\'eatoire par:

$$pv[T(Y)]=\rm E\{ I_{T(Y^*)>T(Y)}|\sigma(Y)\}$$

 o\`u  $Y^*$ est une variable ind\'ependante de $Y$ mais de m\^eme  distribution.
Les tests de permutation sont bas\'es sur un  conditionnement sur les statistiques d'ordre :  $Y_{(o)}=Y_{(1)},\ldots, Y_{(n)}$.

La   p-valeur de permutation  est:
$$ppv[T(Y)]=\rm E\{ I_{T(Y^*)>T(Y)}|\sigma(Y)\vee \sigma(Y^*_{(o)}=Y_{(o)})\}$$
Supposons que nous puissions repr\'esenter $Y$ par $Y=g(\varepsilon)$
avec  $\varepsilon$  \'echangeable. Une telle repr\'esentation 
 a \'et\'e propos\'e par  Cox et Snell \cite{cox68}. Alors $T(Y)=T[g(\varepsilon)]=S(\varepsilon)$. 
Si $\varepsilon$ \'etait observ\'e on pourrait utiliser la p-valeur de permutation :
$$pv_{\varepsilon^*_{(o)}=\varepsilon_{(o)}}[S(\varepsilon)]=\rm E\{ I_{S(\varepsilon^*)>S(\varepsilon)}|\sigma(\varepsilon)\vee \sigma(\varepsilon^*_{(o)}=\varepsilon_{(o)})\}.$$
En g\'en\'eral  $\varepsilon$ n'est pas  observ\'e. Nous  proposons donc de prendre l'esp\'erance:
$$Eppv[T(y)]=\rm E\{pv_{\varepsilon^*_{(o)}=\varepsilon_{(o)}}[S(\varepsilon)]|\sigma(Y)\}.$$
L'esp\'erance peut d\'ependre de param\`etres de nuisance $\gamma\in \Gamma$. Dans ce cas on peut soit les remplacer par les estimateurs du maximum de vraisemblance, soit calculer $\max_{\gamma\in \Gamma} Eppv(\gamma)$. 
 Cette approche est adapt\'ee \`a un mod\`ele de r\'egression logistique.

\section{Introduction}
Permutations tests can be useful as distribution-free tests and also have  exact size  (as opposed to the asymptotic validity of most conventional tests).
However the use of permutation tests in  regression problems has been limited because valid permutation tests obtain only if the observations are exchangeable  under the null hypothesis. A vector $Y$ has an exchangeable distribution if $PY$ has the same distribution as $Y$, for any permutation matrix $P$.
 If we consider a test statistic $T(Y)$, a permutation test is obtained, if $Y$ is exchangeable, by conditioning on the order statistics $Y_{(o)}=\{Y_{(1)},\ldots,Y_{(n)}\}$  \cite{kalb78}. The assumption  of exchangeability, although a little less stringent than the assumption of identically independently distributed (i.i.d.) observations, is still quite restrictive, and does not hold for instance in regression problems.

The has been many applications of permutation tests; a particularly interesting permutation test was proposed by Mantel \cite{mantel67}. Permutation tests are often based on score tests. For some theory about permutation tests see \cite{com03} and for score tests see \cite{com97} and \cite{dry}.

In this paper we propose  a new approach, called expected permutation p-value (Eppv), based on permuting an unobserved exchangeable variable.
 Section 2  presents permutation versions of score tests in generalized linear models. In sectiin 3  some theory about p-values, permutation and conditioning is developed and the Eppv are presented.  This approach is then applied to the logistic regression model in section 4. Section 5 presents a short simulation. An illustration with real data is given ins ectiion 6 which concludes.

\section{Permutation score tests}

Consider a sample of independent random variables $Y_i$, $i=1,\ldots,n$, and assume a generalized linear model; the contribution of observation $i$ to the likelihood  is:
$$
f(Y_i;\theta_i,\eta)= \exp \left\{\eta^{-1}\left[\theta_i Y_i-b(\theta_i)\right]+c(Y_i,\eta)\right\} 
$$
with $E(Y_i)=b'(\theta_i)=\mu _i$ and $\theta_i=Z^i\beta$ where $Z^i=(z_1^i, \ldots, z^i_p)$
 is a row vector of explanatory variables (considered here as deterministic) and $\beta$ is a $p\times 1$ vector of regression coefficients; here $\eta$ denotes the dispersion parameter. Then the score equation obtained by equating to zero the derivative of the loglikelihood $L$ relatively to $\beta$ is 
$Z^T\hat R=0$,
where $Z$ is the $n\times p$ matrix of explanatory variables $z_j^i$ , and $\hat R=(\hat R_1,\ldots,\hat R_n)^T$ is the vector of residuals  $\hat R_i=Y_i-\mu_i(\hat \beta)$. Thus the estimated residuals are orthogonal to the  space of explanatory variables.

 If we consider an explanatory variable indexed by $p+1$, the model becomes $\theta_i=Z^i\beta+z^i_{p+1} \beta_{p+1}$. Lets us denote the parameters $\gamma=(\eta, \beta, \beta_{p+1})$. The score statistic for testing $H_0$: ``$\beta_{p+1}=0$'' has the  linear form:
\begin{equation}S(Y)={\partial L \over \partial \ \beta_{p+1}}(\beta_{p+1}=0)=z_{p+1}^T\hat R,\label{scorestat}\end{equation}
where $z_{p+1}^T=(z^1_{p+1},\ldots,z^n_{p+1})$ is the vector of values for explanatory variable $p+1$ and  $\hat R$ is the vector of residuals in the model not including variable $p+1$. 


A test for $H_0$: ``$\beta_{p+1}=0$'' may be based on the asymptotic distribution of $n^{-1/2}S(Y)$. Let us call  $\phi(Y)$ the critical function of the test ($\phi(Y)=1$: $H_0$ rejected, $\phi(Y)=0$: $H_0$ not rejected); except in simple cases it is not possible to construct exact tests, that is with ${\rm E}_{\gamma} [\phi(Y)]=\alpha$, $\gamma \in \omega$, where $\omega$ is the subset of the parameter space corresponding to $H_0$.  For small sample sizes the difference between the nominal and true Type I error rates may be large. In regression models it is tempting to try to construct tests based on  permutation 
of the residuals in the score statistics \cite{schmoyer94}. Fisher exact test can be shown to be a  permutation of the residuals in a score test, in a case  where the observations are exchangeable under the null hypothesis.  However, generally as soon as there is one explanatory variable under the null hypothesis, neither $Y$ nor $\hat R$ are exchangeable; hence, permutation tests cannot be constructed \cite{com03}.

\section{Some theory about p-values, permutation and conditioning}

\subsection{p-values}

Consider a test $\phi(Y)$ based on a statistic $T(Y)$.
We examine the case where the decision to reject $H_0$ is taken if $T(Y) \ge c_{\alpha}$,
$c_{\alpha}$ being chosen such ${\rm E}_{\gamma} [\phi(Y)]=\alpha$.
A definition of the  p-value which allows to consider it as a random variable (and hence to study its properties) is

$$pv[T(Y)]=\rm E_{\gamma}[ I_{T(Y^*)\ge T(Y)}|\sigma(Y)]$$

where $Y^*$ is a random variable independent from $Y$ but with the same distribution and $\sigma (Y)$ is the sigma-algebra generated by $Y$. See \cite{will} for properties of the conditional expectation.
We can construct a size $\alpha$ test by rejecting $H_0$
if $pv[T(Y)] \le \alpha$, that is: $\phi(Y)=I_{pv[T(Y)] \le \alpha}$.

\subsection{Conditional p-values}
We may define a p-value conditional on ${\cal C}$, where ${\cal C}\subset \sigma(Y,Y^*)$  as:
$$pv_{\cal C}[T(Y)]=\rm E_{\gamma}[ I_{T(Y^*)\ge T(Y)}|\sigma(Y)\vee {\cal C}\}.$$
Conditional tests can be constructed as $\phi(Y)=I_{pv_{{\cal C}}[T(Y)] \le \alpha}$.
We have ${\rm E}_{\gamma} [\phi(Y)|{\cal C}]=\alpha$; it follows that we also have ${\rm E}_{\gamma} [\phi(Y)]=\alpha$. That is, marginally the test has size $\alpha$, but the critical regions (and the power) depend on ${\cal C}$. The conditional approach has been advocated for two different situations \cite{lehmann86}. 

The first arises if we have a sufficient statistic $C$ for the family of measure ${\cal P}^Y=\{P_{\gamma}, {\gamma}\in \omega\}$, where $\omega=H\cap K$, the frontier between the sets representing the null (H) and the alternative (K) hypotheses. If ${\cal C}$ is the sigma-algebra generated by $C$, then $pv_{\cal C}[T(Y)]$ no longer depends on $\gamma$, so that we obtain a similar test, ${\rm E}_{\gamma} [\phi(Y)]=\alpha$, $\gamma \in \omega$.  Such a test is said to have the Neyman structure relatively to  $C$.
As an example consider the case where we observe variables $Y_i$, $i=1,\ldots,n$ which are i.i.d. under the family of measures ${\cal P}^Y=\{P_{\gamma}, {\gamma}\in \omega\}$. 
Then the order statistic.
 $Y_{(o)}=Y_{(1)},\ldots, Y_{(n)}$ is sufficient for $\gamma$ and if we take ${\cal C}=\sigma(\{Y^*_{(o)}=Y_{(o)}\})$ we obtain a permutation test,
 that is we have ${\rm E}[\phi(Y)|Y_{(o)}]=\alpha$. Due to the discrete character of the conditional distribution of $T(Y)$, it is not possible to achieve ${\rm E}[\phi(Y)|Y_{(o)}]=\alpha$ for all $\alpha$, except by resorting to randomisation; we will neglect this problem in the sequel.

The second situation arises in the presence of ancillary statistics $Z$: here the motivation is to perform the test adapted to the situation fixed by the particular realization of $Z$. We may also consider S-ancillary statistics whose distribution depends on an unknown parameter $\xi$, while the distribution of $Y$ given $Z$ does not depend on $\xi$. While the unconditional p-value depends on both $\gamma$ and $\xi$, the p-value conditional on $Z$ does not depend on $\xi$.
As an example consider the case of a regression model where explanatory variables $Z^i$ are associated to response variables $Y_i$: the regression model specifies the conditional distribution of $Y_i$ given $Z^i$ and depend on $\gamma$, while the marginal distribution depends on $\xi$ only. It is natural to consider tests which are conditional on $Z$; in our formalism, for a test stastic $T(Y,Z)$ we then compute the conditional p-value $pv_{\cal C}[T(Y,Z)]$ with ${\cal C}=\sigma(\{Z^*=Z\})$.

The two situations have in common the fact that there is a reduction of the number of parameters  on which the p-value depends. In the particular case where there is a sufficient statistic for $\gamma$, the p-value does not depend on any parameter. However in complex problems this may not be achieved without loosing too much power. One possibility is to replace   $pv_{\cal C}[T(Y);\gamma]$ by  $pv_{\cal C}[T(Y);\hat \gamma]$, where $\hat \gamma$ is an estimator of $\gamma$. We would like to a have a procedure such that 
$|pv_{\cal C}[T(Y);\hat \gamma]-pv_{\cal C}[T(Y);\gamma]|$ is as small as possible. Choosing large ${\cal C}$ may help to reduce the variance of this random variable. Another way is to apply a minimax argument. If it is known that $\gamma$ belongs to a compact set $\Gamma$, then we may base a test on $\max_{\gamma\in \Gamma} pv_{\cal C}[T(Y);\gamma]$. This leads to a test of size lower or equal to $\alpha$.

\subsection{The expected conditional p-value}

Consider the case where $Y=g(\varepsilon)$, where $g(.)$ is a non-decreasing function; if $g$ is not one-to-one we have $\sigma(Y)\subset \sigma(\varepsilon)$. If we have a statistic $T(Y)$, this defines a statistic $S(\varepsilon)=T(g(Y))$. We may consider the p-value
 $pv_{\cal C}[S(\varepsilon)]=\rm E_{\gamma}[ I_{S(\varepsilon^*)\ge S(\varepsilon)}|\sigma(\varepsilon)\vee {\cal C}\}$, where ${\cal C} \subset \sigma(\varepsilon^*,\varepsilon)$.
Since in general this is not $\sigma(Y)$-measurable, we may consider its expectation 
$Epv_{\cal C}[S(\varepsilon)]=\rm E_{\gamma}[pv_{\cal C}[S(\varepsilon)]|\sigma(Y)]$. 
A size-$\alpha$ test can be constructed using this expected conditional p-value as usual.

This approach can in particular be connected with the
 Cox-Snell family which represents $Y$ as
$Y=g(\varepsilon)$, 
where $\varepsilon$ is exchangeable. Such a representation 
was proposed by Cox and Snell \cite{cox68} to define residuals.
If $\varepsilon$ were observed a permutation test could be constructed by conditioning on the order statistic of $\varepsilon$.
It is appealing thus to use an expected conditional p-value choosing 
${\cal C}=\sigma(\varepsilon^*_{(o)}=\varepsilon_{(o)})$. Such a p-value will be called expected permutation p-value (Eppv). 

 Numerically this method is easy to implement: 
draw at random $\varepsilon^*$ from the distribution of $\varepsilon$ conditional on $Y$; compute the permutation p-value; take the  mean of the p-values for a sufficient number of drawings. However the distribution of $\varepsilon$ conditional on $Y$ may depend on parameters that may have to be estimated (see sections 3.2 and 4).

\section{Applications of the Eppv approach to the logistic model}

A logistic regression model is specified by:
$\Pr (Y_i=1)=\pi_i$; logit$(\pi_i)=z^i\beta$. 
It can be depicted in terms of latent i.i.d. variables $\varepsilon_i$ having a uniform distribution on [0,1]:
$$Y_i=I_{\varepsilon_i \le \pi_i}$$

A score test for $H_0:$ ''$\beta_{p+1}=0$''  is
 $T(Y)=S(\varepsilon)=z^T_{p+1}(I_{\varepsilon\le \pi}-\pi)$
with obvious vectorial notation. For a permutation test only the first part $z^T_{p+1}I_{\varepsilon\le \pi}$ is needed. However, because $\sum_i I_{\varepsilon_i\le \pi_i}$ is not constant under permutation of $\varepsilon$, the test is not invariant for a change of origin of $z$:
 there is a need to center one of the two vectors involved in this scalar product, a concept also related to that of 
``clean'' form as in \cite{com03}. Thus the proposed statistic is $T(Y)=S(\varepsilon)=\sum_iz^i_{p+1}(I_{\varepsilon_i\le \pi_i}-n^{-1}\sum_i I_{\varepsilon_i\le \pi_i})=\sum_i (z_{p+1}^i-\bar z_{p+1})I_{\varepsilon_i\le \pi_i})$ (where $\bar z_{p+1}$ is the mean of $z_{p+1}^i$), which is invariant.

For computing  the Eppv  we draw $\varepsilon$ from its conditional distribution which is
\begin{itemize}
\item  $\varepsilon_i \sim U[0, \pi_i]$ if $Y_i=1$;
\item  $\varepsilon_i \sim U[ \pi_i,1]$ if $Y_i=0$.
\end{itemize}

 If the $\pi_i$ are known, an exact permutation test follows.
 In practice one may replace $\pi_i$ by an estimator $\hat \pi_i$, the maximum likelihood estimator of $\pi_i$ under $H_0$, leading to an approximate test. It is conjectured that the type I error probability is $\alpha+ O_p(n^{-1/2})$, similar as when using the asymptotic distribution of the standardized score statistic. However for small sample size the Eppv approach may have better performance because of the non-standard conditioning.
Another possibility is to apply the minimax approach. Consider the case $p=1$ and it is known that $\beta_1\in [a,b]$. One can find $\max_{\beta_1\in [a,b]} Eppv(\beta_1)$ and this leads to a test with  type I error probability lower or equal to $\alpha$. In practice the maximum can be found numerically.

It is interesting to note that when there is no explanatory variable under the null hypothesis, the Eppv test reduces to Fischer's exact test; this happens because for all $i$, $\hat \pi_i=\bar Y$ so that permuting $\varepsilon$ is identical to permuting $Y$.


\section{Simulation study}

We have simulated a Logistic regression model given by:
$$ logit (\pi_i)=\beta_0+\beta_1 z^i _1+ \beta_2 z^i_2$$
with $\beta_0=0$; $\beta_1=1$;
$z_1^i=w_1^i-1$;  $z_2^i=(w_2^i-1)(z_1^i)^d$,
 where $w_1^i$ and $w_2^i$ are independent with  exponential distributions.
 The values  $d=0$, where $z_1^i$ and $z_2^i$ were independent, and $d=1$ and $d=-1$, producing two different cases of non-linear dependencies between $z_1^i$ and $z_2^i$, were tried.
Samples of sizes 30 and 15 were generated from this model.
The problem was: testing $H_0:$ ``$\beta_2=0$'' at size $\alpha=0.05$.
The empirical sizes (for $\beta_2=0$) and powers (for $\beta_2=1$ for $n=30$ and $\beta_2=2$ for $n=15$) of
the   likelihood ratio (LR) test, the  Wald test, a score test based on  permutation of residuals (PR) and the Eppv test have been 
estimated by simulation using 10000 replicates. 
We have also tried a Bootstrap test: among several possibilities we  have chosen the one which seemed the most natural that is a non-parametric bootstrap of  the Wald test; the guidelines given in \cite{hal}, that is resampling  $(\beta_2^*-\hat \beta_2)/\sigma ^*_{\beta_2}$ (where $\beta_2^*$ is the maximum likelihood estimate of $\beta_2$ for a resample and  $\sigma ^*_{\beta_2}$ is the estimated standard deviation of $\beta_2^*$), have been applied; this time-consuming test (using 499 resamples) has been studied on  only 1000 replicates.
For simplicity, for all the tests, only marginal probabilities were estimated, that is we regenerated 
the $z_1^i$ and $z_2^i$ at each replicate.

The results appear in Table 1 (with $\beta_2$   simply denoted $\beta$).
It is clear that the Wald test tends to be conservative while the LR test tends to be anti-conservative. These behaviours are more marked for $n=15$ than for $n=30$. The tests based on permutation better respect the size of the tests with a tendency to conservative for $d=1$; the Eppv test has a better stability than permutation of residuals. The bootstrap Wald test is not really practical for $n=15$ because many  configurations generated by resampling are too particular and lead to failure of convergence of the algorithm; so the results of this test are not displayed in Table 1.  For $n=30$ it is strongly  anti-conservative: the estimated type I error risks are  0.088,  0.097,  0.14 for $d=0$, $1$ and $-1$ respectively. 

The power of the Eppv test is always higher than that of the Wald test and of the test based on permutation of residuals; it is sometimes lower than that of the likelihood ratio test but the latter is not very reliable in the situations considered. In conclusion when working with small samples and when we can suspect a dependency between the factor studied and the other explanatory variables, the Eppv test seems the most reliable among the tests considered here.

\setlength{\parindent}{5mm}

\begin{table}
\caption{Simulation results based on 10000 replicates of a logistic regression model
comparing the Wald test, the likelihood ratio test (LR), the test based on permutation of residuals (PR) and the Eppv test; the theoretical size of the tests is 0.05.}

\vspace{10mm}
\begin {center}

\begin{tabular}{cc|cccc|}

  &  & Wald & LR & PR & Eppv   \\ \hline
 ${\bf n=30}$ & & & & & \\
 &$d=0$ & 0.044 & 0.063 & 0.051& 0.052\\
$\beta=0$ &$d=1$ & 0.025 &0.069 &0.015 & 0.025\\
 (Type I error)&$d=-1$ &0.020 &0.080 &0.062 &0.046 \\ \hline
 &$d=0$ &0.45 &0.53 &0.47 &0.48\\
$\beta=1$ &$d=1$& 0.17 &0.31 &0.14 &0.17 \\
 (Power)&$d=-1$ &0.81 &0.91 & 0.85&0.88 \\ \hline
 ${\bf n=15}$ & & & & & \\
 &$d=0$ &0.020 &0.072 &0.049 &0.049\\
$\beta=0$ &$d=1$ &0.009 &0.094 & 0.018&0.020 \\
(Type I error) &$d=-1$ &0.007 &0.094 &0.066 &0.041\\ \hline
 &$d=0$ &0.22 &0.52 &0.57 & 0.58\\
$\beta=2$ &$d=1$ &0.10 &0.41 &0.19 &0.22\\
(Power) &$d=-1$ &0.16 &0.65 &0.75 &0.79\\ \hline

 \end{tabular}
\end{center}

\end{table}

\section{Illustration on real data}

Even in a large study very small numbers may occur in some categories of 
the sample which are of interest. The small problem treated here for illustration
is taken from a real study on the effect of wine consumption of the risk of developing
dementia \cite{orgo}. In this study, 2273 non-demented subjects were followed up during three years.
Subjects were classified according to their wine  consumption as: no drinkers, 
mild  drinkers moderate or heavy drinkers. During the follow-up 99 cases of dementia developed. Potentially important confounding factors were age, gender and educational level (here coded as a binary variable: no primary diploma vs primary diploma or above).
Globally it appeared from a logistic regression analysis that moderate wine consumption was a protective factor against dementia. However if we try to analyze the data separately 
by gender (which is legitimate because both the curse of dementia and drinking habits are different among genders) very small numbers occur. In particular, there were 28  dementia cases among 
811 non-drinking women and 0 cases among 44 moderate or heavy drinking women. With such figures,
a logistic regression with wine consumption as an explanatory variable fails to converge so that it is not possible to use a Wald test and a likelihood ratio test is probably not very reliable. For one-sided alternative, Fisher's exact test gave a p-value equal to 0.21 and when adjusting on age and educational level, we obtained p-values equal to 0.18 and 0.13
with the PR and Eppv tests respectively; on the basis of these data, taking into account possible confounding factors, the hypothesis that consumption of wine has no effect on risk of dementia among women cannot be rejected.

In conclusion the Eppv approach extends permutation tests ideas to 
complex problems. Bootstrap was also in part motivated by such an extension but unlike bootstrap, the Eppv approach keeps the idea of conditioning on  the order statistic of an exchangeable vector.


\begin{thebibliography}{99}

\bibitem{com03} Commenges, D. (2003), Transformations which preserve exchangeability
 with application to permutation tests, {\em Journal of Nonparametric Statistics} {\bf 59}, 171-195.

\bibitem{com97}  Commenges, D. and Jacqmin--Gadda, H. (1997), Generalized score test of homogeneity based on correlated random effect models, {\em J Roy. Statist. Soc. B} {\bf 59}, 157-171.

\bibitem{cox68} Cox, D. R. and Snell, E. J. (1968), A general definition of residuals, {\em J Roy. Statist. Soc. B} {\bf 30}, 248-75.
\bibitem{dry} Drylewicz, J., Commenges, D. and Thi\'ebaut (2010), Score tests for exploring complex models: application to HIV dynamics model, {\em Biometrical Journal}, {\bf 52}, 10-21.

\bibitem{hal} Hall, P. and Wilson, S.R. (1991), Two guidelines for bootstrap hypothesis testing. {\em Biometrics}, {\bf 47}, 757-762.


\bibitem{kalb78} Kalbfleisch, J.D. (1978),
Likelihood methods and nonparametric tests.
{\em J. Amer. Statist. Assoc.} {\bf 73}, 167-170 .


\bibitem{lehmann86} Lehmann, E.L. (1986), {\em Testing Statistical Hypotheseses} 
New-York: Wiley. 

\bibitem{mantel67} Mantel, N. (1967), The detection of disease clustering and a generalized regression approach, {\em Cancer Research} {\bf 27} Part 1, 209-220.

\bibitem{orgo} Orgogozo, JM., Dartigues, JF., Lafont, S., Letenneur, L., {Commenges D.}, Salamon R., Renaud, S. and Breteler, M. (1997), ``Wine consumption and dementia in the elderly: a prospective community study in the Bordeaux area,'' {\em Revue Neurologique} 153, 185-192.

\bibitem{schmoyer94} Schmoyer, R. L. (1994), Permutation tests for correlation in regression errors. {\em Journal of the American Statistical Association} {\bf  89}, 1507-1516.


\bibitem{will} Williams, D. (1991), {\em Probability with martingales}, Cambridge: Cambridge University Press.

\end{thebibliography}
 \end{document}